\documentclass[prx,amsmath,amssymb, notitlepage, twocolumn,
nofootinbib,
superscriptaddress,
longbibliography
]{revtex4-2}

\usepackage{amsmath}
\usepackage{tabularx,graphicx}
\usepackage{epstopdf}
\usepackage{graphicx}
\usepackage{latexsym}
\usepackage{amssymb}
\usepackage{amsmath}
\usepackage{color, colortbl}
\usepackage{psfrag}
\usepackage{bbm}
\usepackage{bm}
\usepackage{titlesec}
\usepackage{dsfont}
\usepackage{feynmp}
\usepackage{slashed}
\usepackage{multirow}
\usepackage{url}

\newcommand{\beq}{\begin{eqnarray}}
	\newcommand{\eeq}{\end{eqnarray}}

\newcommand{\bsp}{\begin{split}}
	\newcommand{\esp}{\end{split}}

\usepackage{color}
\definecolor{darkblue}{rgb}{0.,0.,0.4}
\definecolor{darkred}{rgb}{0.5,0.,0.}
\definecolor{BlueViolet}{RGB}{138,43,226}
\definecolor{SkyBlue}{RGB}{30,144,255}
\definecolor{DarkGreen}{RGB}{0,100,0}
\usepackage[pdftex,colorlinks=true,linkcolor=darkblue,citecolor=blue,urlcolor=darkred]{hyperref}
\usepackage[normalem]{ulem}

\begin{document}
	\title{Emergent anomaly of Fermi surfaces: a simple derivation from Weyl fermions}
	
	\author{Ruochen Ma}
	\affiliation{Perimeter Institute for Theoretical Physics, Waterloo, Ontario, Canada N2L 2Y5}
	
	\author{Chong Wang}
	\affiliation{Perimeter Institute for Theoretical Physics, Waterloo, Ontario, Canada N2L 2Y5}

\begin{abstract}

The Fermi liquid has an emergent low-energy symmetry that corresponds to charge conservation at each momentum on the Ferm surface. Recently Else, Thorngren and Senthil (ETS) argued that the emergent symmetry had a specific t'Hooft anomaly\cite{ETS}. We give a simple derivation of the ETS anomaly using the chiral anomaly of Weyl fermions.

\end{abstract}

\maketitle

The Landau Fermi liquid\cite{Landau1956}, along with certain types of non-Fermi liquids\cite{Lee2018review}, is characterized by a large emergent symmetry: the charge at each momentum on the Fermi surface is conserved at low energy. Recently, Else, Thorngren and Senthil (ETS) pointed out\cite{ETS} that this large emergent symmetry, mathematically known as the loop group $L^{d-1}U(1)$ where $d$ is the space dimension, has a t'Hooft anomaly of the Chern-Simons form
\beq
\label{ETS}
\frac{i}{(d+1)!(2\pi)^d}\int \mathcal{A}(d\mathcal{A})^d,
\eeq
where the products in the integrand are the wedge product. The probe gauge field $\mathcal{A}$ not only lives in the real spacetime coordinates, but also in the $(d-1)$ dimensional momentum space on the Fermi surface -- this is the manifestation of the charge conservation at each Fermi surface momentum, so the probe gauge field can couple to each momentum separately. As a t'Hooft anomaly, Eq.~\eqref{ETS} lives in one extra dimension, so the integration in total has $(d+1)+(d-1)+1=2d+1$ dimensions. Many known properties of Fermi liquids (and even certain non-Fermi liquids) are nicely unified under this anomaly, including Luttinger theorem, quantum oscillation and anomalous Hall conductivity.

The ETS anomaly Eq.~\eqref{ETS} was proposed\cite{ETS} based on the phase space dimensionality (the Chern-Simons term being essentially the only known form) and the fact that the anomaly reproduces several known properties of Fermi liquids. In this note we give a simple direct derivation of the ETS anomaly.

For simplicity we derive the anomaly for the $(2+1)d$ Fermi surface (the surface itself has one dimension), but our derivation is quite readily generalized to higher dimensions. 

We start from a single Weyl fermion in $(3+1)d$, coupled to an ordinary $U(1)$ gauge field $A$:
\beq
H=\sum_{i=x,y,z}v_i\psi^{\dagger}\sigma_i(i\partial_i-A_i)\psi,
\eeq
where $v_i>0$ is the velocity of the Weyl fermion in the $i$-direction. This theory has the famous chiral anomaly\cite{A,BJ}, described by a Chern-Simons term in $(3+1)+1=5$ dimensions:
\beq
\label{WeylChiral}
\frac{i}{6(2\pi)^2}\int AdAdA.
\eeq
Crucially, this anomaly does not change under various deformations of the theory. In particular, consider turning on a uniform magnetic field in $\hat{z}$ direction $B\hat{z}$. The important feature at low energy is the chiral Landau level, which contains many ($N=\int dxdy B/2\pi$) branches of chiral fermions moving in $\hat{z}$ direction. Now take the Landau gauge $A_x=0, A_y=Bx$, then the momentum $k_y$ is a good quantum number that can be used to label the chiral fermions. At low energy we obtain a \textit{one-dimensional} Fermi surface labeled by $k_y$, with a chiral fermion moving in $\hat{z}$ direction for each momentum $k_y$. Assuming periodic boundary condition in the $\hat{x}$ direction with linear size $L_x$, the momentum $k_y$ takes value in $[0,BL_x)$ with $k_y\sim k_y+BL_x$. We can then interpret $k_y$ as parametrizing a circle, and the Fermi surface is indistinguishable from a conventional one when observed at low energy. Namely, all we have done is to find a slightly unconventional UV regularization of a conventional Fermi surface.

This one-dimensional Fermi surface should of course inherit the original chiral anomaly of the Weyl fermion Eq.~\eqref{WeylChiral}. But what does the $x$ direction in Eq.~\eqref{WeylChiral} mean for the Fermi surface (which has fermions labeled by $k_y$, moving in $\hat{z}$ but no explicit $x$ dependence)? To understand this, we take the single Landau level limit $v_{x}\to \infty$ and completely project the physics to the chiral Landau levels\footnote{Of course in practice we do not need the strict limit of $v_{x}\to\infty$. All that is needed is that all other energy scales (temperature, frequency and interaction strength) are well below the Landau gap $\sim Bv_xv_y$.}. In this limit we can take the standard substitution $x\to k_y/B$, $\partial_x\to B\partial_{k_y}$ and by gauge invariance $A_x\to BA_{k_y}$. The chiral anomaly then becomes exactly the ETS anomaly, where the integration in the $x$ coordinate in Eq.~\eqref{WeylChiral} becomes that of the momentum coordinate in Eq.~\eqref{ETS}. 

As pointed out in Ref.~\cite{ETS}, the $A_{k_y}$ gauge field can be interpreted as the Berry connection on the Fermi surface and contributes to the Hall conductance\cite{Haldane2004}. We can verify this connection explicitly as follows. Consider a configuration $A_x=f(x)\ne 0$, $A_y=Bx$, the equation of motion can still be solved by chiral fermions labeled by $k_y$, with the same number of degeneracy on the lowest Landau level. We then expect the Fermi surface Berry phase to correspond to the magnetic flux through the $x$-circle: $\Phi_B=\oint_x A_x$. The Hall conductance $\sigma_{yz}$ can be measured from the Berry phase of the adiabatic evolution $\phi_{-2\pi}^y\phi_{-2\pi}^z\phi_{2\pi}^y\phi_{2\pi}^z$, where $\phi_{\pm 2\pi}^\alpha$ denotes an adiabatic $\pm 2\pi$ flux threading through the $\alpha$-circle. The effect of $\phi_{\pm2\pi}^z$ is to increase (decrease) the chemical potential in $k_z$ by one minimum unit, namely to increase (decrease) one filled Landau level in $xy$ plane. The Berry phase of the evolution $\phi_{-2\pi}^y\phi_{-2\pi}^z\phi_{2\pi}^y\phi_{2\pi}^z$ is therefore equal to the Berry phase of the simple flux-threading $\phi_{2\pi}^y$ for a single filled Landau level. The latter is simply given by $\oint_x A_x=\Phi_B$, exactly what we anticipated.    

We also note that when going from the Weyl chiral anomaly to the ETS anomaly, we could subtract off the ``background" magnetic field by making the substitution $A_y\to A_y+Bx\sim A_y+k_y$. This results in an additional piece that can be simply interpreted as the sum of the chiral anomalies of $(1+1)d$ chiral fermions. This simple sum of $(1+1)d$ chiral anomaly is always present and is not included in the ETS anomaly.

To recap, what we have done is to find a UV regularization -- a Weyl fermion under a magnetic field -- of a Fermi surface, from which the IR anomaly follows simply from the UV anomaly. The advantage of this regularizations is that the conserved charge at each point on the Fermi surface (the generator of $LU(1)$) is local (in $x$), in sharp contrast to usual settings in which the conserved charge at a specific momentum $k_y$ is highly non-local in real space. The locality of the generators allow them to be coupled to a probe gauge field, which is fundamentally why the information of the $LU(1)$ ETS anomaly can be encoded in the UV chiral anomaly.

We end with several additional remarks:
\begin{enumerate}

\item The $(3+1)d$ Weyl fermion also comes with a mixed $U(1)$-gravity anomaly. This anomaly becomes a mixed $LU(1)$-gravity anomaly for the Fermi surface. This has a familiar interpretation: the Berry phase on the Fermi surface gives a contribution to the thermal Hall conductance of the metal\cite{Haldane2004}.

\item It is rather straightforward to generalize our argument to higher dimensions: for the Weyl fermion, we can increase the space dimension by $2$ (recall that Weyl fermion exists in even spacetime dimensions), and put another magnetic flux density in the two new dimensions. We will again have chiral Landau levels (ultimately guaranteed by the chiral anomaly of the Weyl fermion). Repeating our earlier argument for the two new dimensions, it is easy to see that we have increased the Fermi surface dimension by one and converted the other dimension into momentum. We can repeat this procedure arbitrarily to access higher dimensions, and the ETS anomaly Eq.~\eqref{ETS} follows simply from the chiral anomaly of a single Weyl fermion in dimension $D=2d$.

\item There is one curious feature when the Weyl fermion lives in dimension $D=4n+2$, where a pure gravitational anomaly exists\cite{GAnomaly}. One natural question is whether this pure gravitational anomaly has any implication for the Fermi surface (in spacetime dimension $(2n+2)$). However, since the Fermi surface is quite trivial when the $LU(1)$ is completely broken (the Fermi surface can easily be gapped then), it seems unlikely that an analogue of pure gravitational anomaly will play a serious role for the Fermi surface. We leave a detailed analysis of this aspect to future work.

\end{enumerate}

\begin{acknowledgements}

We thank T. Senthil and Liujun Zou for illuminating discussions. We also thank Anton Burkov, Alexander Hickey and Xuzhe Ying for discussions and for a related collaboration. RM and CW acknowledge support from the Natural Sciences and Engineering Research Council of Canada (NSERC) through a Discovery Grant. Research at Perimeter Institute is supported in part by the Government of Canada through the Department of Innovation, Science and Industry Canada and by the Province of Ontario through the Ministry of Colleges and Universities.

\end{acknowledgements}

\bibliography{Ref.bib}

\end{document}